\documentclass{elsart_special}
\journal{Nuclear Instruments and Methods in Physics Research A}

\usepackage{epsfig}

\usepackage{amssymb}

\begin{document}
\begin{flushright}
CLNS 01/1754
\end{flushright}

\begin{frontmatter}



%
\title{Active cooling control of the CLEO detector using a hydrocarbon
coolant farm}


\author[Cornell]{A. Warburton\corauthref{cor1}},
\corauth[cor1]{Corresponding author.}
\ead{AndreasWarburton@mailaps.org}
\author[Purdue]{K. Arndt}, 
\author[Cornell]{C. Bebek\thanksref{LBNL}}, 
\thanks[LBNL]{Current address: Lawrence Berkeley National Laboratory,
Berkeley, California  94720, USA.}
\author[Cornell]{J. Cherwinka\thanksref{Wisconsin}},
\thanks[Wisconsin]{Current address: Focused Research, Inc., Middleton,
Wisconsin  53562, USA.}
\author[WSU]{D. Cinabro}, 
\author[Purdue]{J. Fast\thanksref{FNAL}},
\thanks[FNAL]{Current address: Fermi National Accelerator Laboratory,
Batavia, Illinois  60510, USA.} 
\author[Cornell]{B. Gittelman},
\author[Cornell]{Seung J. Lee}, 
\author[WSU]{S. McGee},
\author[Cornell]{M. Palmer},
\author[WSU]{L. Perera\thanksref{Rutgers}},
\thanks[Rutgers]{Current address: Rutgers University, Piscataway, New Jersey
  08855, USA.}
\author[Minnesota]{A. Smith},
\author[Purdue]{D. Tournear\thanksref{Stanford}},
\thanks[Stanford]{Current address: Stanford University, Stanford, California
  94305, USA.}
and
\author[Cornell]{C. Ward\thanksref{Intel}}.
\thanks[Intel]{Current address: Intel Corporation,
Hillsboro, Oregon  97124, USA.}

\address[Cornell]{Wilson Synchrotron, Laboratory of Nuclear
Studies, Cornell University, Ithaca, New York  14853, USA}
\address[Purdue]{Purdue University, West Lafayette, Indiana  47907, USA}
\address[WSU]{Wayne State University, Detroit, Michigan  48202, USA}
\address[Minnesota]{University of Minnesota, Minneapolis, Minnesota  55455,
USA}

\begin{abstract}
We describe a novel approach to particle-detector cooling in which a
modular farm of active coolant-control platforms provides independent
and regulated heat removal from four recently upgraded subsystems of
the CLEO detector: the ring-imaging \v{C}erenkov detector, the drift
chamber, the silicon vertex detector, and the beryllium beam pipe.  We
report on several aspects of the system: the suitability of using the
aliphatic-hydrocarbon solvent PF\texttrademark-200IG as a
heat-transfer fluid, the sensor elements and the mechanical design of
the farm platforms, a control system that is founded upon a commercial
programmable logic controller employed in industrial process-control
applications, and a diagnostic system based on virtual
instrumentation.  We summarize the system's performance and point out
the potential application of the design to future high-energy physics
apparatus.
\end{abstract}

\end{frontmatter}


\section{CLEO experiment}
\label{sect:cleo3}

The CLEO experiment, located on the Cornell Electron Storage Ring
(CESR), consists of a general-purpose particle physics
detector~\cite{CLEOII,SVX} used to reconstruct the products of
symmetric $e^+e^-$ collisions with centre-of-mass energies up to
approximately 10~GeV, corresponding to the $\Upsilon(4S)$ resonance.
Luminosity-increasing improvements in the CESR optics, which required
the placement of superconducting quadrupole magnets nearer to the
$e^+e^-$ interaction region, precipitated upgrades to several inner
subsystems of the detector prior to the CLEO III datataking run.

The main outer components of the CLEO detector are a CsI
electromagnetic calorimeter, a 1.5~T solenoidal superconducting
magnet, and muon detectors.  The upgraded inner subsystems consist of,
in order of decreasing radius with respect to the $e^+e^-$ beam line,
the following four devices: a 230$\,$400 channel ring-imaging
\v{C}erenkov (RICH) detector for charged-hadron
identification~\cite{RICH}; a new central drift chamber~\cite{DR3},
with 9796 cells arranged in 47 layers, for charged-particle tracking;
a four-layer double-sided silicon microstrip vertex detector~\cite{Si3}
with 125$\,$000 readout channels for precise tracking and decay
vertexing; and a thinly walled beryllium beam pipe~\cite{BP3} to
minimize tracking uncertainties due to multiple-scattering effects.

An active coolant-control farm, which is the focus of this article,
was used to provide heat removal and mechanical stability (via
temperature control) for the above four upgraded inner CLEO
subsystems.  In the case of the RICH detector, the principal heat
sources were chains of Viking~\cite{Viking} front-end signal
processing chips, with a combined power output of $\sim$360~W.  The
drift chamber's heat sources consisted of a total of $\sim$150~W of
power dissipated by pre-amplifiers mounted on each of the two end
plates; temperature stability across the end plates was crucial to
prevent wire breakage.  The silicon detector produced $\sim$488~W of
power from 122 BeO hybrid boards of front-end electronics
chips~\cite{SI3_power}, each dissipating 4~W; spatial tracking
resolution of the silicon layers depended on the alignment precision
of the sensors and therefore the temperature stability.  The Be beam
pipe was itself passive, but required a design heat-removal capability
up to the $0.5 - 1$~kW range due to the possibility of higher-order
mode heating in the CESR $e^+e^-$ collider.

In this article, we describe the approach taken in designing the CLEO
coolant-control farm, studies of the hydrocarbon
PF\texttrademark-200IG and its suitability as a heat-transfer fluid,
mechanical construction of the farm platforms, the sensor elements,
the process-control and diagnostic systems, and the farm's performance
in the CLEO III datataking run.

\section{Design approach}
\label{sect:approach}

\subsection{Generality, scalability, and modularity}

The principal challenge posed in the design of the CLEO cooling system
lay in the requirement that multiple detector subsystems, each with
markedly different plumbing layouts, power loads, pressure tolerances,
and physical locations, were to be actively cooled by a `farm' of
independent coolant-control platforms governed and monitored by
centralized control and diagnostic systems, respectively.  Rather than
design separate custom coolant circuits tailored to the specifications
of each detector subsystem, our approach was to found the system on a
single simple generic design, one that possessed the flexibility to
satisfy the cooling requirements of any of the subsystems without
extensive modification.

The advantages of a scalable generic design were numerous.  Each
system had the same basic set of fittings and sensors, introducing
economies of scale and easing spare-component inventories.  The
modularity of the design was intended to allow for the rapid swapping
of an entire coolant-control platform with a spare unit in case of a
problem.  The offending platform could then be removed {\it en masse}
from the radiation area for diagnosis and repair, without the accrual
of excessive downtime.  Technician training, serviceability, and the
management of on-call experts were simpler in this farm paradigm;
separate cooling experts were not required for different detector
subsystems, and the total number of experts needed was reduced.

Our use of a modular farm of nearly identical cooling platforms lent
itself well to a unified process-control system, which we describe in
Section~\ref{sect:control}.  Scaling of the control electronics to
accommodate additional coolant platforms was designed to be a facile
matter of adding more input/output channels and updating the control
logic.  All electronic sensors, including those not used for
process-control variables, were read out by the control electronics.
A separate diagnostic system, described in
Section~\ref{sect:diagnostics}, therefore had no need to interact with
the farm sensors directly; instead, all the diagnostic information was
acquired through the control infrastructure.  The diagnostic system
was designed to provide globally accessible, minute-by-minute
performance information browsable on the World Wide Web ({\sc www}).

Beneficial to the farm concept was a simple, rapid, and independent
channel of communication between the coolant-control electronics and
the detector subsystems proper.  Instead of using a subsystem-based
temperature reading, which would have required significant
customization on each platform, the cooling platforms used the
temperatures of their coolant supply as process variables, thereby
precluding a need for the detector subsystems to communicate with any
aspect of the cooling system.  Communication in the other direction,
however, namely from the process-control electronics to the detector
subsystems, was possible in the form of interlocks that the cooling
system could breach in the event of serious performance problems.

Each cooling platform was designed to maintain a specific,
user-defined, fixed coolant-supply set-point temperature at
approximately a constant flow rate of heat-transfer fluid, with an
active feedback system that automatically compensated for changes in
subsystem thermal power load, ambient temperature and humidity
changes, and variations in heat-sink water temperatures and flow
rates.  Supply set-point temperatures were remotely selectable, and
were designed to remain at most within $\pm 0.3$~K of the requested
temperature.  All the farm platforms were designed to deliver flow
rates up to approximately 23~L/min and to handle heat loads of up to
approximately 1~kW.  Table~\ref{tab:specs} provides a summary of the
principal operating parameters of a generic CLEO coolant-control
platform design; subsystem-specific requirements are described in
Section~\ref{sect:special} below.

\begin{table}
\caption{Summary of principal design parameters of a generic
CLEO cooling platform.}\label{tab:specs}
\begin{center}
\begin{tabular}{lcc}
\hline
Parameter & Type & Value \\
\hline
Coolant set-point temperature & minimum & 287~K \\
                              & nominal & 292~K \\
                              & maximum & 299~K \\
Cooling capacity & maximum & 1~kW \\
Flow rate, main circuit & maximum & 23~L/min \\
Particulate impurity size & maximum & 75~$\mu$m \\
Platform pressure & minimum & atmospheric \\
                  & maximum & 690~kPa \\
Set-point temperature stability & minimum & $\pm$0.3~K \\
                                & maximum & $\pm$0.1~K \\
Subsystem pressure drop & maximum & $\sim$200~kPa \\
\hline
\end{tabular}
\end{center}
\end{table}

\subsection{Special subsystem requirements}
\label{sect:special}

Although the farm model was based upon identical platforms, some
customization was nevertheless necessary to accommodate the specific
needs of each of the new subsystems, described in
Section~\ref{sect:cleo3}, serviced by the coolant-control farm.

In the upgraded CLEO beam pipe, the thickness of beryllium separating
the cooling channels from the inner accelerator vacuum volume and the
outer atmosphere was 737~$\mu$m and 229~$\mu$m,
respectively~\cite{BP3}.  The differential pressure limit on the
cooling-channel walls was therefore required not to exceed
$\sim$203~kPa, necessitating the implementation of overpressure
safeguards in the beampipe coolant-control platform (refer to
Section~\ref{sect:mech}).

Large pipe gauges in the plumbing leading to the drift-chamber cooling
lines required a pump with a higher head rating to achieve adequate
flow rates (refer to Section~\ref{sect:mech}).  The resultant
increased flow capacity of the drift-chamber coolant-control platform
allowed for this module to be used to provide temperature control to
other passive devices (see Section~\ref{sect:conclusion}).  Due to the
drift chamber's large size and mechanical sensitivity to uneven
temperature distributions on the end plates, a dedicated interlock
(described in Section~\ref{sect:control}) was required to prevent
situations in which only a subset of the pre-amplifier electronics,
mounted near the end plates, was powered while one set of global
cooling parameters was applied to the entire heat-transfer area.

A similar motivation drove the customization in the control of
heat-transfer fluid inside the RICH detector, for which there was a
design necessity to be able to vary, in response to real-time
conditions, the relative coolant flow rates through five azimuthally
distinct regions of the detector.  Rather than modify the RICH
coolant-control platform itself, the main module was left generic and a
separate ``active-manifold'' platform was designed to accommodate the
furcation of a single global process-control loop into five
individually controlled flow circuits.  In Sections~\ref{sect:mech}
and \ref{sect:sensors}, aspects of the mechanical design and the
special array of RICH coolant-temperature sensors required to
construct and operate, respectively, the RICH active-manifold platform
are discussed.

The upgraded silicon vertex detector was installed into the CLEO
interaction region $\sim$4~months later in the commissioning period
than the beam pipe, drift-chamber, and RICH subsystems; however,
during this time, the silicon detector underwent electronics testing
at a location $\sim$60~m remote from the CLEO experiment.  The control
system design therefore needed to be able to provide uninterrupted
interlock management and active coolant control of the installed
systems while concurrently and flexibly supporting the testing of the
silicon detector at the remote location.

\section{Hydrocarbon heat-transfer fluids}
\label{sect:coolant}

The liquid coolant employed to remove heat from the detector
subsystems consisted of PF\texttrademark-200IG, an industrial solvent
and degreaser manufactured by P-T Technologies, Inc.~\cite{PTT}.  We
know of no other applications that exploit the thermodynamic and flow
characteristics of this substance in lieu of its chemical-contaminant
solvency properties.  Accordingly, we investigated several attributes
of the PF\texttrademark-200IG product, including its applicability to
the cooling requirements of CLEO detector subsystems.

\subsection{Coolant requirements}

The hygroscopic nature of the cesium iodide in the CLEO barrel and
endcap calorimeters~\cite{CLEOII} demanded that a heat-transfer fluid
be employed in which CsI crystals were insoluble.  Initially, the
halocarbon compound 1,1,2-Trichloro-1,2,2-Trifluoroethane
(Freon\texttrademark\footnote{``Freon\texttrademark'' is du Pont's
registered trademark for its fluorocarbon compounds~\cite{Freon}.} TF
R-113 or Arcton\texttrademark\footnote{``Arcton\texttrademark'' is a
registered trademark of Imperial Chemical Industries Limited.} 113)
was used as a coolant, but its utility was limited by its
ozone-depleting potential, its reactivity with beryllium, a propensity
to leak without straightforward detection, cost, and a tendency to
modify non-metal fittings such as ceramic pump seals, elastomers, and
plastic sensor housings.

A second major consideration for a heat-transfer fluid to replace
Freon\texttrademark~was therefore the requirement that it be
compatible with the materials in each coolant circuit.  In particular,
this was a concern for the cooling channels in the thinly walled
beryllium beam pipe, where corrosion avoidance was a necessity;
Section~\ref{sect:Becompat} describes a study that addressed this
compatibility.

In addition to the needed compatibility with CsI and Be, a suitable
coolant was required to have thermodynamic and flow characteristics
comparable to those of water.  The heat-transfer fluid of choice was
to be environmentally benign, be reasonably safe to handle, be
non-corrosive to several different materials, and have preferably low
electrical conductivity and negligible amounts of non-volatile residue
remaining after evaporation.\footnote{In the event of a leak, high
resistivity and complete evaporation were deemed necessary to minimize
damage to electronics and detector components.}

\subsection{Properties of PF\texttrademark-200IG}

The substance PF\texttrademark-200IG is a non-flammable, electrically
non-conductive, 100\% volatile, aliphatic-hydrocarbon solvent with no
ozone-depleting potential.  It is a complex combination of normal
paraffins (Alkanes) consisting of a straight chain of non-aromatic
saturated hydrocarbons having carbon numbers in the range of C-5
through C-20~\cite{CAS-PF200}.  At ambient temperatures and pressures,
PF\texttrademark-200IG is a colourless liquid that has a faint
petroleum odour and is insoluble in water.  The notation ``200IG''
identifies the product as of industrial grade, with a 200~$^\circ$F
(366~K) tag-closed-cup flash point.

Table~\ref{tab:props} compares several properties of
PF\texttrademark-200IG with liquid water and Freon\texttrademark~TF.
We note that PF\texttrademark-200IG has $\sim$27\% the thermal
conductivity and $\sim$55\% the specific heat capacity of water.
Taking into account the lower density and higher kinematic viscosity
(refer to Table~\ref{tab:props}), PF\texttrademark-200IG has
approximately a factor of two worse heat-transfer capability as
compared to water, for a given pumped plumbing circuit.

We studied the PF\texttrademark-200IG radiation length using 20~keV
photons, finding it to be $\sim$52~cm ($\sim$44\% longer than that for
water).  Particularly in the case of the beryllium beam pipe, where a
cooling channel surrounded the vacuum chamber throughout the
interaction region, the longer radiation length advantageously reduced
the degradation of tracking resolution due to multiple scattering.

\begin{table}
\caption{Properties of PF\texttrademark-200IG contrasted with those of
liquid water and Freon\texttrademark.}\label{tab:props}
\begin{center}
\renewcommand{\arraystretch}{2.4}
\begin{tabular}{lccccc}
\hline
Property & PF\texttrademark-200IG & Water & Freon\texttrademark~R-113
& Units & Temp. \\
\hline

\begin{minipage}{0.9in}
Auto-ignition\\temperature
\end{minipage}
& 483~\cite{PTT} & $-$ & $-$ & [K] & $-$ \\

\begin{minipage}{0.9in}
Boiling-point\\temperature
\end{minipage}
& $466 - 516$~\cite{PTT} & 373 & 321~\cite{Freon} & [K] & $-$ \\

\begin{minipage}{0.9in}
CAS registry\\number
\end{minipage}
& 64771-72-8 & 7732-18-5 & 76-13-1 & [$-$] & $-$ \\

\begin{minipage}{0.9in}
Chemical\\constitution
\end{minipage}
&
\begin{minipage}{0.9in}
\begin{center}
Paraffins,\\C5-20~\cite{CAS-PF200}
\end{center}
\end{minipage}
&H$_2$O & CCl$_2$F-CClF$_2$ & [$-$] & $-$ \\

Flash point & 366~\cite{PTT} & $-$ & 623~\cite{Freon} & [K] & $-$ \\

\begin{minipage}{0.9in}
Heat of\\vapourization
\end{minipage}
& 247~\cite{PTT} & 2257~\cite{HCP} & 147~\cite{Freon}
& [kJ/kg] &
\begin{minipage}{0.5in}
\begin{center}
boiling\\point
\end{center}
\end{minipage}
\\

\begin{minipage}{0.9in}
Kinematic\\viscosity
\end{minipage}
& 2.41~\cite{PTT} & 0.89~\cite{HCP} & 0.44~\cite{Freon} & [cSt] &
$\sim$296~K \\

\begin{minipage}{0.9in}
Radiation\\length
\end{minipage}
& 0.52 & 0.361~\cite{PDG} & 0.15~\cite{PDG} & [m] & ambient \\

\begin{minipage}{0.9in}
Specific\\gravity
\end{minipage}
& 0.79~\cite{PTT} & 1 & 1.56~\cite{Freon} & [$-$] & ambient \\

\begin{minipage}{0.9in}
Specific heat\\capacity
\end{minipage}
& 2303~\cite{PTT} & 4186~\cite{HCP} & 892~\cite{Freon} &
[J/kg$\cdot$K] & $\sim$292~K \\

\begin{minipage}{0.9in}
Thermal\\conductivity
\end{minipage}
& 0.159 & 0.5984~\cite{HCP} & 0.074~\cite{Freon} &
[W/K$\cdot$m] & $\sim$294~K \\

\begin{minipage}{0.9in}
Vapour\\pressure
\end{minipage}
& 0.041~\cite{PTT} & 6.75~\cite{HCP} & 62~\cite{Freon} & [kPa] & 311~K
\\

\hline
\end{tabular}
\end{center}
\end{table}

\subsection{Compatibility of PF\texttrademark-200IG with beryllium and
other materials}
\label{sect:Becompat}

The PF\texttrademark-200IG solvent was described by the manufacturer
as non-corrosive to metals such as aluminium, copper, magnesium, and
stainless steel~\cite{PTT}.  Motivated by the type of metal used to
construct the CLEO beam pipe, we undertook a study to examine the
compatibility of PF\texttrademark-200IG with beryllium.

Three beryllium plates with dimensions 5.3~cm $\times$ 2.5~cm $\times$
0.3~cm were each coated on one side with a corrosion resistant primer
(BR127) and immersed in volumes of de-ionized water,
PF\texttrademark-200IG, and air (the control sample), respectively.
Following a period of 3 months at ambient temperature, pressure,
and light exposure, there were no visibly discernible changes; the
samples were subsequently placed near CESR where they absorbed a
radiation dose of $\sim$25~krad.

\begin{figure}
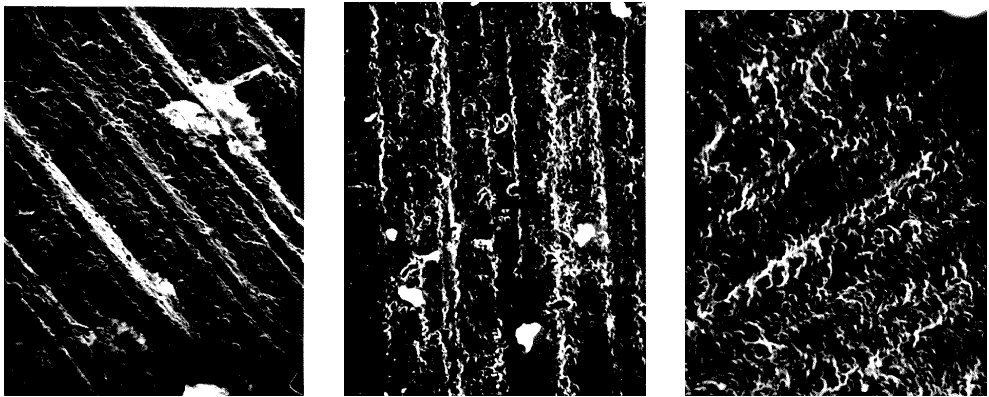

\epsfxsize=4.cm
\hspace{0.1in}
\epsfbox{be_air.epsi}
\epsfxsize=4.cm
\hspace{0.1in}
\epsfbox{be_water.epsi}
\epsfxsize=4.cm
\hspace{0.1in}
\epsfbox{be_pf200.epsi}

\caption{Scanning electron micrographs of three beryllium surfaces
(magnified $\sim$250 times) following contact with, from left to
right, (a) air, (b) de-ionized H$_2$O, and (c) PF\texttrademark-200IG
after a period of over 3 months and a radiation dose of
$\sim$25~krad.  Note that all three samples show milling grooves with
comparable periodicity.}
\label{fig:sem}
\end{figure}

With still no visually apparent deterioration in the exposed or
primer-coated sides of the beryllium plates, we used a scanning
electron microscope to examine the sample surfaces, micrographs of
which are depicted in Figure~\ref{fig:sem}.  We observed that the
sample that was in contact with PF\texttrademark-200IG
(Figure~\ref{fig:sem}(c)) exhibited less surface modification than the
sample that was in contact with H$_2$O (Figure~\ref{fig:sem}(b)), as
compared to the control sample (Figure~\ref{fig:sem}(a)).  Based on
these observations, we concluded that PF\texttrademark-200IG was at
least as compatible with beryllium as de-ionized water.

In addition to our studies with beryllium, we tested the compatibility
of PF\texttrademark-200IG with other materials: cesium iodide, Buna
and Viton\texttrademark~elastomers, Push-Lok\textregistered~rubber
hose, brass and stainless-steel fittings, copper, aluminium, plastic,
and both polypropylene and nylon tubing.  Our tests consisted of
recording the masses and dimensions of the material samples and
immersing them in containers of PF\texttrademark-200IG under ambient
conditions for a period of $\sim$8 weeks, whereupon we measured the
mass changes and the fractions of linear swell.  For each of the
materials tested, the observed changes in mass and size were
negligible.

\section{Mechanical construction}
\label{sect:mech}

The mechanical design of the coolant-control platforms was driven
primarily by the following criteria: reliability and serviceability;
modularity; elevation and footprint; mobility; and ease of access to
gauges, valves, filters, and reservoirs.  The limited space in the
``pit'' beneath the CLEO detector, an approximately 16~m$^2$ area of
$\sim$97~cm high crawl space, dictated a footprint of 76~cm $\times$
76~cm and an elevation of $\sim$87~cm for each of the coolant-control
platforms, including the RICH active-manifold platform.  Each member
of the farm was constructed on a rubber-footed skid of 0.6~cm thick
aluminium, enabling a degree of mobility and easy access to platform
facilities.

\begin{figure}
\epsfxsize=14.cm
\epsfbox{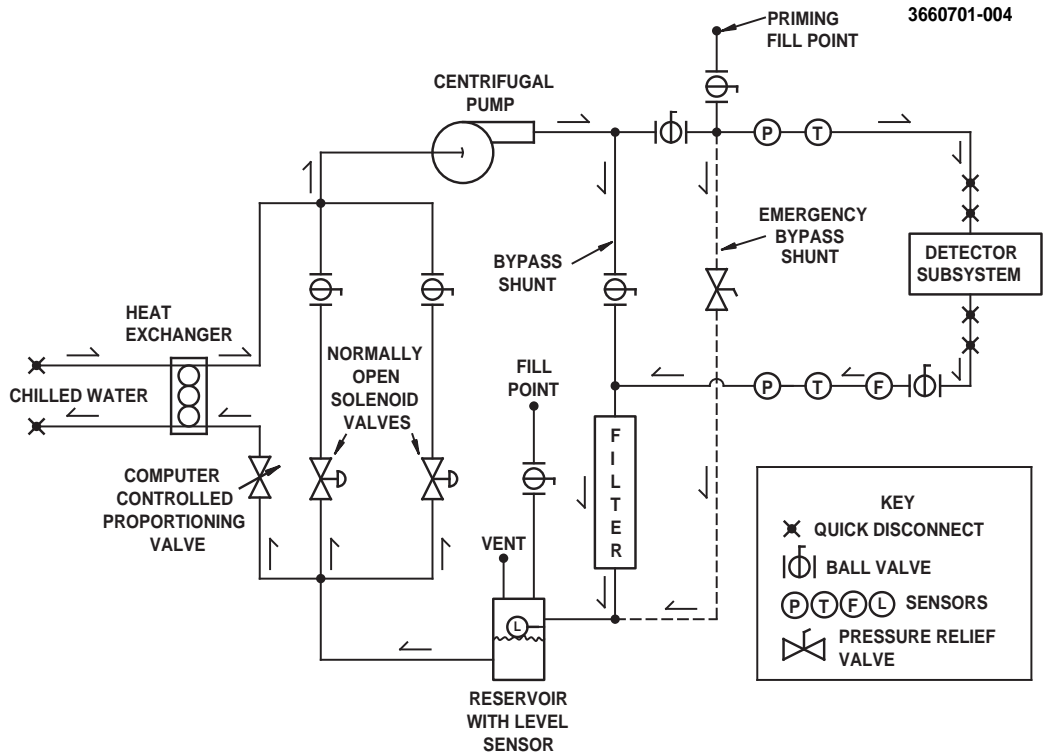}
\caption{Schematic of a hydrocarbon coolant circuit.}
\label{fig:platform_schematic}
\end{figure}

\begin{figure}
\centerline{\epsfig{figure=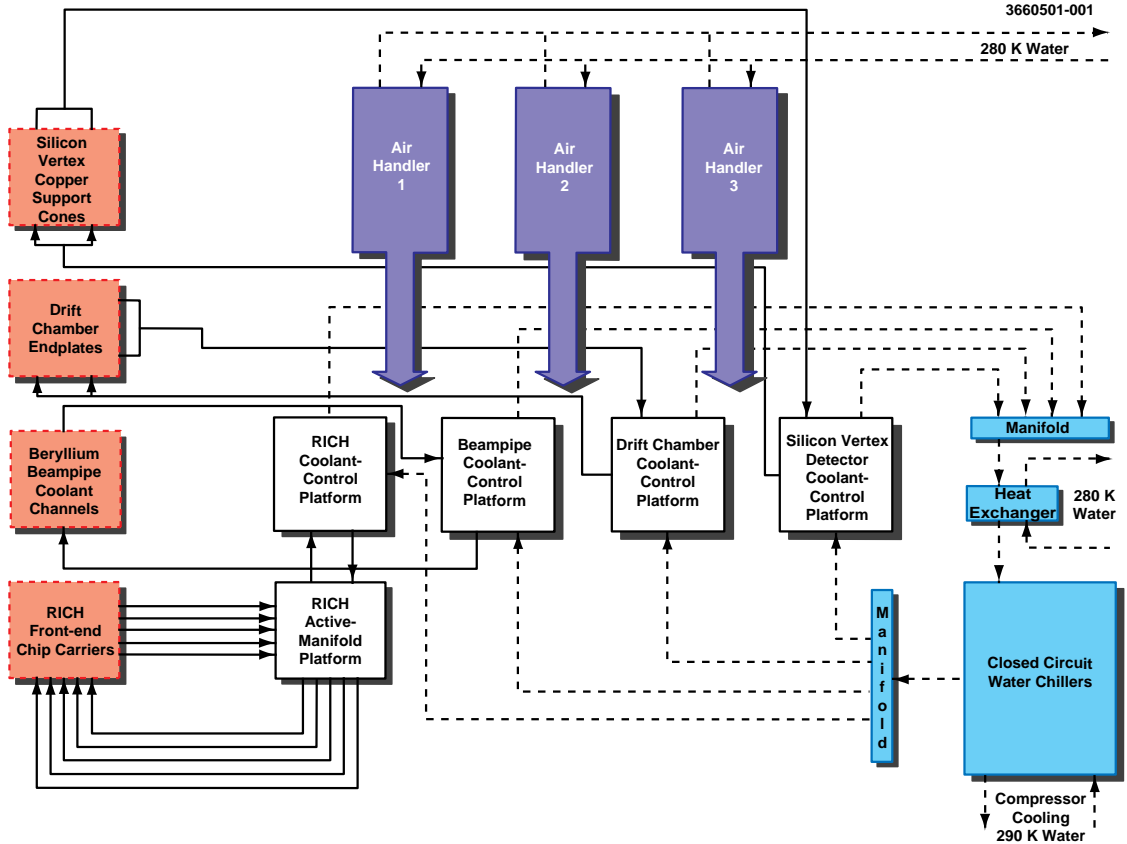,width=4.5in,angle=-90}}
\caption{Schematic layout of the CLEO coolant-control farm, showing
flow circuits of the three cooling fluids: air (wide arrows), liquid
water (dashed lines), and liquid hydrocarbon (solid lines, each
representing an individually regulated circuit).  The bifurcations in
the supply and return lines to and from the silicon vertex detector
and the drift chamber represent the division of coolant flow between
the electron and positron sides of the CLEO detector.}
\label{fig:farm_schematic}
\end{figure}

Figure~\ref{fig:platform_schematic} depicts a schematic of the coolant
flow circuit on board one of the platforms in the farm.  Upon leaving
the reservoir due to suction from the pump, PF\texttrademark-200IG
coolant reached the pump inlet by way of either a branch through the
hot side of a heat exchanger or via one of two bypass shunts.  The
cold side of the heat exchanger, the primary heat sink for the
platform consisting of a 20-brazed-plate Cetetherm (Cetetherm AB,
Ronneby, Sweden) honeycomb unit, was connected to a closed water
circuit driven by a chiller system (refer to
Figure~\ref{fig:farm_schematic}).  The amount of flow through the heat
exchanger was regulated by a proportioning valve powered by a
computer-controlled step motor; the fraction that the valve was open
constituted the control variable in the feedback system.  Unregulated
12~Vdc power supplies were mounted on each platform to power the
proportioning-valve step motors.

The two heat-exchanger shunt branches each contained a ball
valve\footnote{Ball valves were preferred over gate, globe, and needle
valves because of their larger proportionality constants describing
the achievable flow rates per unit of pressure drop at a given
temperature.} and a normally-open solenoid valve.  One design goal of
the bypass system was to guarantee that there be flow to the detector
subsystem at all times, assuming that both solenoid valves could never
be simultaneously closed while the proportioning valve was 0\% open.
Dual bypass branches were implemented in order to expand the dynamic
range of the system using binary flow logic.  For a given fixed
fractional flow rate in the bypass branches, large variations in the
subsystem's heat load meant that the proportioning valve alone would
not provide enough compensation to maintain a stable set-point
temperature.  Using solenoid valves to switch the two bypass branches,
each with a fixed ball-valve setting, resulted in four discrete bypass
flow configurations.  With ball-valve settings appropriately chosen,
the proportioning valve could provide full analog coverage for
continuous flow changes intermediate to the four binary combinations
of the two solenoid valves.  We note that an alternative to this
scheme to maximize the dynamic range would be simply to use a second
proportioning valve in the heat-exchanger bypass shunt; we did not
adopt this approach for reasons of reliability and economy.

Once through the outlet port of the pump, the PF\texttrademark-200IG
could either bypass back through the filter and into the reservoir or
depart the platform for transport to the detector subsystem, as
indicated in Figure~\ref{fig:platform_schematic}.  The global rate of
PF\texttrademark-200IG flow leaving the platform was configured by
partially closing the ball valve on the bypass branch leading back to
the reservoir.  Maximizing the fraction of flow through this bypass
branch greatly assisted the feedback process by pre-cooling the
temperature of the reservoir contents to near the set-point value.

\begin{figure}
\epsfxsize=16.cm
\hspace{-1.0cm}
\epsfbox{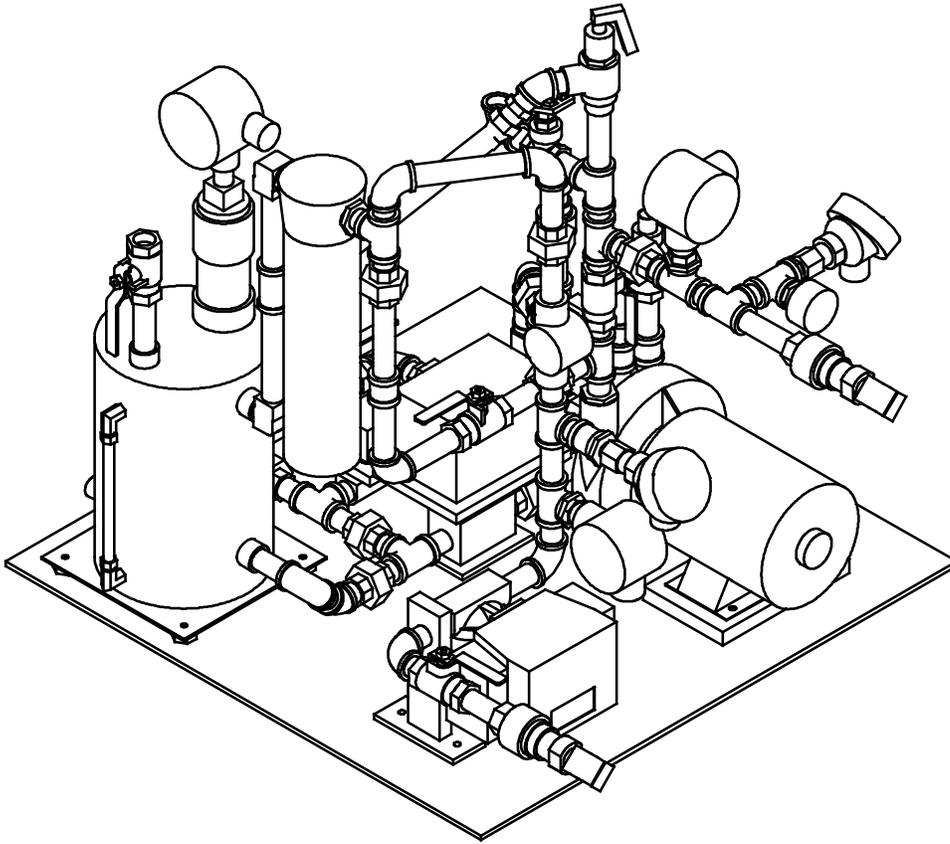}
\caption{Isometric assembly drawing of a coolant-control platform
module, with the reservoir and the pump motor discernible in the left
and right corners, respectively.}
\label{fig:platform_drawing}
\end{figure}

The plumbing on the coolant-control platforms consisted primarily of
brass 3/4'' NPT threaded pipe fittings connected using approximately
79 nipples (sealed with Loctite PST\textregistered~567) per platform.
Each platform also had $\sim$7 ball valves, $\sim$17 elbows, $\sim$6
couplings, $\sim$14 tees, and several bushings.  Reservoirs were
constructed from stainless steel, and the filter housings\footnote{Our
prototype platforms used polypropylene filter housings; although
sprayed with conductive anti-static paint and fitted with ground
straps, significant charge build-up due to the moving
PF\texttrademark-200IG dielectric separating the filter (also
polypropylene) from the housing still resulted in some arcing.  We
therefore caution against the use of non-metallic housings and
recommend that each coolant-control platform be carefully grounded.},
also stainless steel, contained 75~$\mu$m polypropylene filters.  The
self-priming centrifugal pumps had stainless-steel casings, 750~W
(2.2~kW for the drift-chamber platform) single-phase electric motors,
and a 34~m (45~m for the drift-chamber platform) water head rating.
Our determinations of these pump-head specifications took into account
the flow requirements through the detector subsystems, the reduced
density of PF\texttrademark-200IG (refer to Table~\ref{tab:props}),
the diameters and elevations of the plumbing runs, and the need for
adequate flow in the bypass branches for optimum temperature control
of the heat-transfer fluid.  Throughout the plumbing of each platform,
12 stainless-steel unions were used to aid in the servicing of
different components.  An additional emergency bypass shunt containing
a 690~kPa pressure relief valve linked the pump outlet port with the
reservoir in the event of an overpressure situation.  Refer to
Figure~\ref{fig:platform_drawing} for an assembly drawing of a typical
coolant-control platform.

At the maximum 23~L/min flow rates required by the design, the vapour
pressure of PF\texttrademark-200IG (refer to Table~\ref{tab:props})
was low enough to ensure that the cavitation
number\footnote{Cavitation number is a dimensionless quantity
expressed as a quotient of the dynamic pressure, which is proportional
to the density and the square of the flow rate, and the difference
between the static and vapour pressures.} was well in excess of the
incipient cavitation value.  No pressurization of the coolant circuits
beyond ambient was therefore deemed necessary to avoid cavitation,
permitting us to vent any of the reservoirs to the atmosphere during
operation.  In practice, we only vented the reservoir in the case of
the beryllium beampipe cooling platform, where the differential
pressure limit on the thin walls of the beryllium cooling channels was
required not to exceed $\sim$203~kPa (refer to
Section~\ref{sect:special}), a criterion that we also explicitly
imposed on the system pressure near the outlet port of the pump on the
platform proper.  In lieu of the fixed (690~kPa) pressure relief
valve, we installed an adjustable unit with the range 103 $-$ 165~kPa.
As an extra precaution, the beampipe coolant-control platform also had
a graphite rupture disc rated at 207~kPa.  During prototype testing,
we observed that this disc would rupture at what appeared to be lower
than the rated pressure, as indicated by a glycerin-damped analog
visual gauge.  We ascribed this to transient pressure pulses from the
pump's impeller and relief-valve oscillations; we subsequently
installed a water-hammer suppressor in an effort to minimize the
effect of these pulses.

Figure~\ref{fig:farm_schematic} depicts the main elements of the
coolant-control farm, located in the pit beneath the CLEO detector,
and the flow of cooling fluids (air, liquid water, and
PF\texttrademark-200IG hydrocarbon) between them.  Wherever possible,
brass quick-disconnect fittings were used to link up the liquid
connections, which were insulated with AP Armaflex\texttrademark.  A
closed-circuit water chiller system provided water near a temperature
of 282~K to the cold sides of the heat exchangers residing on each of
the coolant-control platforms (refer to
Figure~\ref{fig:platform_schematic}).  In order to assist the water
chillers, a heat exchanger to 280~K building water was inserted to
pre-cool the returning coolant.  Three air handlers, their compressor
coils cooled using 280~K building water, maintained a flow of cool dry
air across the farm platforms.  Also shown in
Figure~\ref{fig:farm_schematic} is the RICH active-manifold platform,
a dedicated module that consisted of two one-to-five manifolds, the
supply manifold outfitted with five computer-controlled proportioning
valves and the return manifold instrumented with five flow meters and
transmitters.

\section{Sensor elements}
\label{sect:sensors}

On or near the coolant-control platforms, the relatively harsh
environment due to dust, water, hydrocarbons, EMI from pump and fan
motors, and vibration demanded a design that made use of industrial
sensor technologies.  In order to minimize the amount of delicate
electronics in the CLEO pit, the analog-to-digital conversion of
sensor signals took place in the control-system crate, described in
Section~\ref{sect:control}.  Standard industrial 4$-$20~mA transmitter
technology was used to condition and send analog sensor signals from
the farm platforms to the control system.  Current transmitters had
the advantages of greater noise immunity and an ability to send analog
signals over relatively long distances.  In addition, the remote
sensors in the CLEO pit could be powered (usually using $\sim$3.5~mA
of excitation current) and read out using a single shielded two-wire
current loop, with no local power supply requirements.  Because of the
possibility for rigourous environmental conditions in the CLEO pit,
all sensors in the design were required to have sealed weatherproof
enclosures that fulfilled the specifications of the NEMA-4X standard.

As indicated in Figure~\ref{fig:platform_schematic}, each
coolant-control platform was instrumented with two temperature
sensors, one supply and one return, to provide a measurement of the
temperature rise due to the combination of subsystem power load,
frictional heating of the moving coolant, and ambient heat transfer.
The coolant supply temperature measurement was particularly critical
since it served as the process variable in each of the main feedback
control systems.  This required the use of temperature sensors that
were accurate and stable to less than $\pm$0.1~K with low noise and
good linearity characteristics.  We investigated four temperature
measurement technologies: thermocouples, resistance temperature
detectors (RTDs), thermistors, and solid-state integrated circuits.
Thermistors were too unreliable and nonlinear.  We used solid-state
devices, based on the AD592 (Analog Devices, Inc.)  integrated
circuit, in a prototype coolant-control platform, finding them to be
attractive due to their relatively good linearity and well-defined
characteristics, but difficult to insert into the coolant flow stream
reliably and necessitating the design and production of a custom,
weatherproof, diode-protected, 4$-$20~mA transmitter circuit (using,
{\it e.g.}, the AD693 device).  We concluded that the AD592
integrated-circuit sensors were better suited for measuring
temperatures at the surfaces of solid objects rather than inside
flowing fluids.  We disfavoured a thermocouple solution because of the
costs of the cold-junction compensation and linearization circuitry
needed to achieve the desired accuracy.  Our final design used
tip-sensitive 100~$\Omega$ (industry standard IEC 751) platinum RTDs
manufactured by Minco Products, Inc.; once combined with some
relatively simple off-the-shelf linearization and 4$-$20~mA
transmitter circuitry, RTDs were more linear, accurate, and sensitive
than thermocouple implementations of comparable cost.  A weatherproof
316 stainless-steel thermowell assembly safely housed the RTD and the
front-end electronics, allowed for the effective insertion of the RTD
directly into the flow of coolant, and permitted the replacement of
the platinum element without having to breach the volume of
heat-transfer fluid.

In addition to the pairs of temperature sensors instrumenting each of
the farm platforms (as shown in Figure~\ref{fig:platform_schematic}),
the RICH system had a dedicated array of 32 tip-sensitive 100~$\Omega$
platinum RTDs distributed azimuthally and on both the electron and
positron sides of the detector in coolant-return manifolds located
near the RICH modules.  For these RTDs, `hockey-puck' style $4-20$~mA
transmitters were mounted together in a single array located atop the
CLEO detector; shielded wire leads, 18.3~m in length, connected the
RTDs to these dedicated transmitters, which were calibrated to
compensate for the net lead resistance.  A custom-built multiplexer
circuit facilitated addressed read-out of any one of the 32
temperature transmitters at a time.

Each coolant-control platform had visual pressure and flow-rate gauges
to aid in manual adjustments to the flow configuration.  In addition,
every platform transmitted two pressure measurements, {\it i.e.}, the
pressure drop through the CLEO subsystem and its plumbing network
(refer to Figure~\ref{fig:platform_schematic}), to the control crate
using two 0$-$689~kPa Series 634E pressure sensors from Dwyer
Instruments, Inc.  Flow rates of heat-transfer fluid were transmitted
from density-compensated vane-type flow sensors manufactured by
Universal Flow Monitors, Inc.  Level transmitters in the coolant
reservoirs were manufactured by Omega Engineering, Inc., and consisted
of a magnet that was mounted inside a stainless-steel float that
tracked up and down an insertion stem while setting a series of reed
switches in a voltage divider resistor network.  The accuracies of the
pressure, flow, and level sensors and transmitters used were
$\sim$14~kPa, $\sim$1~L/min, and $\sim$1.3~cm, respectively.

\section{Process-control system}
\label{sect:control}

The heart of the coolant-control system consisted of a small logic
controller (SLC) module, a member of the SLC~500\texttrademark~family
of programmable controllers manufactured by Allen-Bradley Company,
Inc., that resided in a dedicated 13-slot chassis with an integrated
1~A (24 Vdc) power supply module.  The SLC~5/04 module used was
capable of up to 4096 inputs plus 4096 outputs and had a memory of
32~K words.  The remainder of the process-control system consisted of
three other module varieties mounted in the chassis: a 32-channel
current-sourcing digital DC output module (1746-OB32), three 4-channel
0$-$20~mA analog output modules (1746-NO4I) with 14-bit resolution,
and eight 4-channel $\pm$20~mA ($\pm$10~Vdc) analog input modules
(1746-NI4) with 16-bit resolution.

The 13-slot Allen-Bradley chassis was mounted inside a crate enclosure
positioned in the experimental hall outside the main radiation area.
Also residing in the enclosure was a 24~Vdc/12~A regulated power
supply used to energize the entire set of $4-20$~mA sensors remotely
deployed both on the coolant farm platforms underneath the CLEO
detector (refer to Figure~\ref{fig:farm_schematic}), $\sim$35~m away,
and in the array of 32 RICH temperature transmitters on top of the
detector, $\sim$20~m away.  Terminal blocks mounted on a DIN rail in
the enclosure served to interconnect the 24~Vdc power supply, the
individual sensors, and the appropriate terminals of the analog input
modules in the Allen-Bradley chassis.  In the case of the 32 RICH
transmitter signals, since the array was multiplexed, only a single
analog input channel was required.  The SLC clocked through the 32
addresses in 0.25~s intervals by using five of the digital output
channels in the 1746-OB32 module to switch the 24~Vdc of the main
sensor power supply.

The SLC processor used a ladder-logic programming language in which
subroutines were organized into ladders, their rungs each acting as
{\sc if-then} conditional statements.  On the left side of every rung
one or more conditions were defined; the corresponding right side
executed one or more actions provided that all the conditions were met
for the given rung.  Editing of the ladder-logic code was achieved
with RSLogix~500\texttrademark~(Rockwell Automation) programming
software running on a networked Intel\texttrademark-based computer
located in a clean computing environment and connected to the
Allen-Bradley control crate by a 30~m RS-232 serial connection.  The
ladder-logic code was compiled with the
RSLogix~500\texttrademark~software and was communicated to the
SLC~5/04 module using a dedicated utility (RSLinx\texttrademark).  In
a similar manner, this serial communication configuration had the
capability to allow online edits, parameter adjustments, and
diagnostic readout of inputs, outputs, and internal memory structures
{\it during} programme execution in the SLC.

The Allen-Bradley SLC control code consisted of $\sim$260 rungs
organized into a main ladder that cycled through a series of calls to
14 subroutine ladders.  For each of the beampipe, drift-chamber, RICH,
and silicon subsystems there were subroutines for platform sensor data
acquisition, interlock decisions and output, and process control.  The
RICH system had two extra subroutines, one to read out the multiplexed
RICH temperature signals and one to perform process-control duties and
flow-sensor data acquisition for the RICH active-manifold platform.

During every cycle of control code execution, an interlock decision
was taken for each of the farm platforms.  The four criteria forming
this decision consisted of a minimum coolant flow rate, minimum and
maximum set-point temperatures, and a minimum level of heat-transfer
fluid in the reservoir (for leak detection).  Supply and return
pressure criteria were not included in the interlock decisions.  For a
given farm platform, if these conditions were all satisfied, the
interlock ladder logic would use a channel in the 1746-OB32 output
module to energize a normally open relay switch mounted to the DIN
rail near the terminal blocks in the chassis enclosure.  The
cooling-interlock relays were connected to a higher-level interlock
crate that could switch off power to the subsystem electronics crates
in the event of a failure.  The control system was designed such that
there would also be an interlock breach if the Allen-Bradley SLC had
any interruption in power.

Specific to the drift-chamber coolant-control platform, an additional
interlock was used to reduce the possibility of cooling the chamber
end plates unevenly, a situation that potentially posed deleterious
consequences to the chamber's mechanical integrity.  If part or all of
the drift-chamber electronics underwent an unexpected loss of power,
the power to the centrifugal pump on the coolant-control platform was
switched off by means of a relay on a 10-minute delay.  In turn, this
would render the minimum-flow-rate criterion unsatisfied, thereby
breaking the cooling interlock and removing power from all of the
drift-chamber electronics crates.

At the core of the process-control logic in each closed feedback loop
was a proportional integral derivative, or PID, instruction tuned to
maintain a desired setting of an input process variable by computing
appropriate real-time adjustments to an output control variable.  For
each coolant-control platform, the process variable consisted of an
input from the platinum RTD sensor measuring the temperature of the
heat-transfer fluid supplied to the detector subsystem; the control
variable consisted of the fractional opening of the heat-exchanger
proportioning valve, as set by the step motor positioned using
4$-$20~mA signals from the 1746-NO4I analog output modules (refer to
Figure~\ref{fig:platform_schematic}).  In order to increase step-motor
life, a deadband of 0.1~K was used in the PID algorithm; in this way,
the control variable was left unchanged once the process variable
passed through the set point and until it was different from the set
point by the deadband amount.


\section{Diagnostic system}
\label{sect:diagnostics}

For reasons of reliability and flexibility, the diagnostic and
process-control systems for the CLEO coolant-control farm were kept
relatively independent through the use of a two-tiered configuration.
Notwithstanding, the same networked computer that was used to upload
the compiled ladder-logic code into the Allen-Bradley SLC, as
described in Section~\ref{sect:control}, served as the run-time
implementation platform for the diagnostic software.

A graphical programming language was used to develop a user interface,
also graphical, to the diagnostic system parameters.  The software was
based on virtual instruments (VIs) in the
LabVIEW\texttrademark~environment provided by National Instruments
Corporation.  During normal farm operation, {\it i.e.}, when the SLC
was in a `standalone' mode and assumed sole control of the farm, the
RS-232 serial connection between the SLC and the networked computer
(refer to Section~\ref{sect:control}) could be used by the
LabVIEW\texttrademark~software to read out regions of the SLC memory
for diagnostic purposes.  Code from proprietary driver and VI
libraries (HighwayVIEW\texttrademark, from SEG, Watertown MA, USA)
transacted the data between the SLC and the
LabVIEW\texttrademark~software.

Although the diagnostic interface primarily read data from the SLC for
the purpose of displaying them to users, it did provide
password-privileged cooling experts with the ability to adjust the
platform set-point temperatures maintained by the control system.
Such set-point changes were effected, again using a
HighwayVIEW\texttrademark~VI, by writing data into SLC memory via the
serial interface.  Users of the local area network could therefore
``window in'' to the LabVIEW\texttrademark~computer to view real-time
diagnostics and, if necessary, make limited adjustments to the control
parameters.

\begin{figure}
\epsfbox{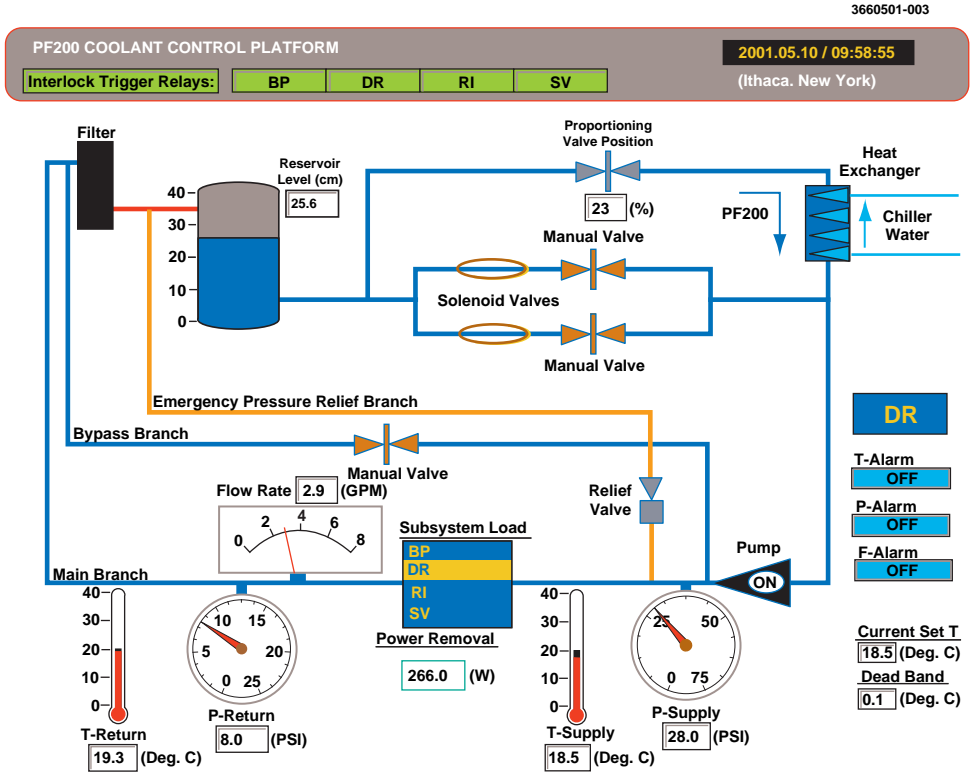}
\caption{Web-based diagnostic panel for real-time viewing of the
principal performance parameters of a given coolant-control platform,
in this example that for the drift-chamber (DR) detector subsystem.
Similar displays describe the other subsystems.  This image was
recorded during the CLEO III datataking run.}
\label{fig:LabVIEWFP_subsystem}
\end{figure}

Diagnostic information was made available to other subsystem experts
and data-acquisition personnel by hypertext transfer-protocol (http)
servers running on the LabVIEW\texttrademark~computer.  For each
platform in the farm, there was an http server capable of delivering
streaming quasi-real-time images of the main
LabVIEW\texttrademark~front panel to a client web browser.  An example
of one of these web-based front-panel displays is given in
Figure~\ref{fig:LabVIEWFP_subsystem}.  Other web accessible displays
included special diagnostics for the RICH temperature-sensor array
(Section~\ref{sect:sensors}), RICH active-manifold parameters
(Section~\ref{sect:mech}), and virtual strip charts giving a graphical
representation of the 12-hour history of some of the more important
system parameters.

In addition, every minute, the LabVIEW\texttrademark~machinery updated
a detailed text-only html file containing an expert diagnostics digest
that listed the status of all farm parameters.  This diagnostic file
was written to public web space and was accessible remotely using the
{\sc www}.  In particular, the digest file format lent itself well to
remote small-screened portable devices like personal digital
assistants using mobile infrastructure software such as that available
from AvantGo, Inc.  Long-term archiving of the coolant-control farm
operating history was achieved by taking half-hourly snapshots of the
expert diagnostics digest file, compressing them, and storing them in
a database.

\section{System performance}
\label{sect:performance}

Maintenance of the farm platforms proved to be minimal, consisting of
weekly visual inspections and annual cleanings and polypropylene
filter replacements.  Some minor coolant leaks, requiring occasional
top ups, were detected by the observation of changes in reservoir
level readings with time.  In the beampipe and silicon-detector
systems, which each had longer runs of Push-Lok\textregistered~rubber
hose connecting the platforms to the subsystems, the
PF\texttrademark-200IG acquired an orange colour, contrary to earlier
compatibility studies (refer to Section~\ref{sect:Becompat}).  We
attributed this discolouration to the dissolution of a powdered
protectant with which the inner surfaces of the
Push-Lok\textregistered~rubber had been treated; although no
degradation of the hose was observed, we recommend using polypropylene
or nylon tubing in future applications.

Operationally, the dynamic ranges of the proportioning valves alone
precluded a need to switch the solenoids in the heat-exchanger bypass
shunts in order to maintain PID control.  With fixed ball-valve
openings in the two shunt branches and a deadband setting of
$\pm$0.1~K in all the PID loops, supply temperatures were asserted
with a stability of $\pm$0.2~K with respect to the set-point values
for the duration of the CLEO III datataking period.  This degree of
control was exploited during CLEO III commissioning in pedestal
sensitivity studies of the CLEO beampipe PIN-diode radiation
monitors~\cite{BP_radmon}, for which the beampipe coolant-control
platform was employed to vary the coolant supply temperature between
293~K and 299~K.

The CLEO coolant-control farm commenced operation in November 1999
with the beginning of the CLEO III commissioning period.  As described
in Section~\ref{sect:special}, prior to March 2000 the silicon farm
platform was deployed remotely and was successfully operated and
monitored by the central farm control and diagnostics systems,
respectively.  On 16 March 2000, a building cooling-water pipe burst,
flooding the CLEO pit and submerging the farm platforms under
$\sim$25~cm of water; all sensors, power supplies, and pumps, after
drying and cleaning, survived the event.  On a few other occasions,
many of the farm's interlocks were exercised in power failures or due
to temperature variations in the 290~K water supply that caused trips
in the chiller compressors (refer to Figure~\ref{fig:farm_schematic}).
The Allen-Bradley SLC~5/04 module (described in
Section~\ref{sect:control}), which was equipped with a battery back-up
system, was rendered immune to brief power glitches and, it was found,
could be relied upon as though it were a firmware device.

The CLEO III detector began taking physics quality data in July 2000
for a run that ended in June 2001, with a scheduled three-week down in
September 2000.  The coolant-control farm induced no CLEO or CESR
downtime in this period, during which a time-integrated luminosity of
$\sim$9~fb$^{-1}$ was accumulated.

\section{Conclusions and outlook}
\label{sect:conclusion}

We have described a novel approach to active particle-detector cooling
that is based upon a farm of modular coolant-control platforms charged
with the hydrocarbon solvent PF\texttrademark-200IG, uniquely used as
a heat-transfer fluid~\cite{PTT}.  During the CLEO III datataking run,
the farm provided reliable cooling support to the RICH detector, the
drift chamber, the silicon vertex detector, and the beryllium beam
pipe, with a temperature stability that exceeded design
specifications.

The CLEO hydrocarbon coolant farm will see continued service in an
upcoming programme of operation to explore the $\Upsilon$ resonances,
charm physics, and quantum chromodynamics
(CESR-c/CLEO-c)~\cite{CLEO-c}.  As mentioned in
Section~\ref{sect:special}, the drift-chamber platform, by virtue of
its greater flow capacity, is being upgraded to provide additional
cooling for passive permanent-magnet (NdFeB) quadrupole
elements~\cite{PM} that are being installed prior to the $\Upsilon$
resonance running period.  Other aspects of the farm will remain the
same.

We have shown that a centrally controlled and monitored farm of
generic active coolant-control platforms, running the liquid
hydrocarbon solvent PF\texttrademark-200IG as a heat-transfer fluid,
can provide independent and regulated heat removal from several
different subsystems in a particle-physics detector.  Aspects of this
design are applicable to future high-energy physics apparatus where
flexibility, minimal maintenance, and the ability to monitor and
operate detector and accelerator systems remotely will be
progressively important.

\section*{Acknowledgements}
We would like to thank K.~Powers and G.~Trutt of the Wilson
Synchrotron Laboratory, Cornell University, for their excellent
technical support.  This work was supported by the U.S. National
Science Foundation, the U.S. Department of Energy, and the Natural
Sciences and Engineering Research Council of Canada.

\end{document}